\documentclass[
reprint,
superscriptaddress,
amsmath,
amssymb,
showkeys,
aps,
prl,
nolongbibliography,
]{revtex4-2}

\usepackage{threeparttable}
\usepackage{setspace}
\usepackage{makecell}
\usepackage{amsmath}
\usepackage{graphicx}
\usepackage{dcolumn}
\usepackage{bm}
\usepackage{hyperref}
\usepackage[mathlines, pagewise]{lineno}
\usepackage{textcomp}
\usepackage{gensymb}
\usepackage{url}
\usepackage{xcolor}
\usepackage{soul}
\usepackage{hhline}
\usepackage[version=3]{mhchem}
\usepackage{multirow}
\usepackage[american]{babel}
\usepackage{textgreek}
\usepackage{romannum}
\usepackage{epstopdf}
\usepackage{subcaption}

\begin{document}

\title{Band Offsets at \texorpdfstring{\textit{\textbeta/\textgamma}-\ce{Ga2O3}}{} Interface}

\author{Huan Liu}
\affiliation{Department of Physics, University of Helsinki, P.O. Box 43, FI-00014, Helsinki, Finland}

\author{Ilja Makkonen}
\affiliation{Department of Physics, University of Helsinki, P.O. Box 43, FI-00014, Helsinki, Finland}

\author{Calliope Bazioti}
\affiliation{Department of Physics and Centre for Materials Science and Nanotechnology, University of Oslo, P.O. Box 1048 Blindern, N-0316 Oslo, Norway}

\author{Junlei Zhao}
\affiliation{Department of Electronic and Electrical Engineering, Southern University of Science and Technology, Shenzhen, 518055, China}

\author{Alexander Azarov}
\affiliation{Department of Physics and Centre for Materials Science and Nanotechnology, University of Oslo, P.O. Box 1048 Blindern, N-0316 Oslo, Norway}

\author{Andrej Kuznetsov}
\affiliation{Department of Physics and Centre for Materials Science and Nanotechnology, University of Oslo, P.O. Box 1048 Blindern, N-0316 Oslo, Norway}

\author{Flyura Djurabekova}
\affiliation{Department of Physics, University of Helsinki, P.O. Box 43, FI-00014, Helsinki, Finland}

\begin{abstract}
Ultrawide bandgap semiconductor gallium oxide (\ce{Ga2O3}) and its polymorphs have recently attracted increasing attention across physics, materials science, and electronics communities. In particular, the self-organized formation of the $\beta / \gamma$-\ce{Ga2O3} double polymorph structures was demonstrated recently [A. Azarov et al., Nat. Commun. 14, 4855 (2023)], paving the way for prospective applications of such structures in electronics. Consequently, determining the conduction band offset in such structures is crucial since it dictates the behavior of conduction electrons at the interface and, consequently, the potential functionality of such interfaces. Thus, in this work, we calculate the band offsets at the $\beta / \gamma$-\ce{Ga2O3} interface using density functional theory in correlation with the data provided by the experimental atomistic interface analysis. Specifically, to unravel the strain state of the $\beta / \gamma$-\ce{Ga2O3} interface, nanoscale strain maps were recorded using high-resolution transmission electron microscopy. In its turn, theoretically, lineup potential and vacuum alignment methods were used to analyze the band offsets, with and without strain, at the $\beta / \gamma$-\ce{Ga2O3} interface. Altogether, the collected results suggest that the band offsets between the $\beta$ and $\gamma$ phases are likely not exceeding a few hundred meV, remaining highly sensitive to the strain state at the interface. At this end, we conclude that even though the formation of a two-dimensional electron gas (2DEG) at the $\beta / \gamma$ interface is theoretically possible, the gradual strain relaxation—if it occurs as a function of the distance from the interface—poses a significant challenge, as it may shift the 2DEG localization or even reduce the overall probability of its formation.
 
\end{abstract}

\maketitle

\section{Introduction}\label{introduction}

Gallium oxide (\ce{Ga2O3}) has garnered considerable interest due to its unique properties, establishing it as a crucial topic in physics, materials science, and electronic engineering. \ce{Ga2O3} boasts an ultra-wide bandgap (4.8-5.3 eV), exceptional thermal stability (melting point $> 1800$ C$^{\circ}$), and high breakdown electric field ($\sim8$ MV/cm), making it an ideal candidate for a diverse array of electronic applications, ranging from high-voltage power electronics~\cite{higashiwaki2016recent, pearton2018review, xue2018overview, zhang2022ultra}, high-frequency devices~\cite{moser2020toward, green2022beta}, solar-blind ultraviolet optoelectronics~\cite{guo2019review, hou2020review, shi2021wide}, high-temperature gas sensors~\cite{SFzhao2021two} to large-scale atomic passivation thin films~\cite{wurdack2021ultrathin, gebert2023passivating}.

\ce{Ga2O3} exhibits several polymorphic phases~\cite{roy1952polymorphism, yoshioka2007structures, SFzhao2023complex}. Monoclinic $\beta$-\ce{Ga2O3} is the most stable phase at room temperature and atmospheric pressure. Other experimentally known metastable phases, such as orthorhombic $\kappa$-phase, corundum $\alpha$-phase, bixbyite $\delta$-phase, and defective spinel $\gamma$-phase can be observed in the form of thin films under high-temperature and high-pressure conditions. The $\kappa$-phase (also referred to as $\epsilon$-phase in a part of literature~\cite{cora2017real, mezzadri2016crystal, cora2020situ}) reveals ferroelectric properties that are favorable for the high-density two-dimensional electron gas in electronic devices. The $\alpha$-phase is particularly interesting for its wider bandgap compared to the $\beta$-phase, making it suitable for high-power and high-frequency electronic devices. The $\gamma$-\ce{Ga2O3} exhibits ultrahigh radiation tolerance~\cite{SFazarov2023universal, SFzhao2024crystallization}, which can be useful for electronics under extreme operating conditions. The $\delta$-phase is less explored in the literature, however, highlighted by its synthesis on a $\beta$-\ce{Fe2O3} buffer layer~\cite{kato2023demonstration}.

Recently it has been shown that the $\beta$-\ce{Ga2O3} transforms into a new polymorph through disorder-induced ordering under ion irradiation~\cite{azarov2022disorder, yoo2022atomic, SFazarov2023universal, huang2023atomica, huang2023atomicb, SFzhao2024crystallization}. Specifically, upon reaching a transition disorder threshold, the metastable $\gamma$-\ce{Ga2O3} phase becomes more favorable than the disordered $\beta$-\ce{Ga2O3}. Moreover, the $\beta$-to-$\gamma$ transformation requires only migration of Ga atoms, while the oxygen sublattice remains intact~\cite{he2024ultrahigh, wang2023size, SFzhao2024crystallization}.
Indeed, the existence of the $\gamma$-phase significantly enhances the radiation tolerance of \ce{Ga2O3} material, retaining crystalline even after exposure to ion irradiation with ultrahigh fluences~\cite{SFazarov2023universal}. This feature is explained by an intriguing combination of the Ga- and O-sublattices properties~\cite{SFazarov2023universal, he2024ultrahigh}. For instance, as shown in Ref.~\cite{he2024ultrahigh}, the O-sublattice tends to efficiently recover face-centered-cubic stacking under radiation conditions. 
Amazingly, an abrupt $\gamma/\beta$-interface forms due to this process, too \cite{SFazarov2023universal, SFzhao2024crystallization, azarov2022disorder}. 
Such double $\gamma/\beta$ polymorph structures separated by an abrupt interface resemble classical hetero-structures, however, without a requirement of changing the chemical composition, motivating further studies of their structural and electronic properties. Indeed, the conduction band minimum (CBM) and valence band maximum (VBM) are predominantly formed out of the Ga-$4s$ and O-$2p$ orbitals respectively~\cite{swallow2020influence}, without any chemically-governed effect~\cite{swallow2021indium}, diverging the band edges in $\gamma$- and $\beta$-phases because of their different symmetries.
Consecutively, the absence of chemical changes in polymorphic heterostructure implies a serious challenge for the experimental band offset determination, which would otherwise be readily performed  
using X-ray photoelectron spectroscopy (XPS), as it was done at the interfaces between $\beta$-\ce{Ga2O3} and other semiconductors, such as Si and Ge~\cite{gibbon2018band, chen2016band}, \ce{In2O3}~\cite{sun2018band}, \ce{Al2O3}~\cite{kamimura2014band}, AlN~\cite{chen2018band}, MgO~\cite{matsuo2018epitaxial}, GaAs~\cite{ji2021determination}, and \ce{LiGa5O8}~\cite{zhang2024experimental}. Indeed, the major obstacle for using XPS to measure band offsets at the $\beta / \gamma$-\ce{Ga2O3} interface is exactly because there are identical atoms on both sides of the interface, indistinguishable in terms of their core-level electron structures. On the other hand, \textit{Ab initio} hybrid-functional density functional theory (DFT) calculation for $\beta / \gamma$-\ce{Ga2O3} interface are possible similarly to that for the interface between $\beta$-\ce{Ga2O3} and \ce{Al2O3}~\cite{lyu2023band}, AlN and GaN~\cite{lyu2020band}, (Al$_x$Ga$_{1-x}$)$_2$O$_3$~\cite{mu2020orientation}. 

Thus, to tackle this problem and to delve into the intriguing properties of the band offset at the $\beta / \gamma$-\ce{Ga2O3} interface, our strategy for this paper is to combine structural $\beta / \gamma$ interface measurements performed by electron microscopy, including strain mapping, with DFT studies. In this way, we endeavor to elucidate the key factors influencing band offset, such as interface structure, stacking orientations, and strain. The strain state is predicted to play a crucial role in determining band offsets, which has also been studied with the geometric phase analysis (GPA). 
As such,  the data collected in this study paves the way for significant improvements in the design and optimization of the next-generation gallium oxide-based devices, contributing to the ongoing evolution of semiconductor technology.

\section{Results}\label{results}

\subsection{Interface structures}\label{interface}

\begin{figure*}
     \centering
     \begin{subfigure}{0.49\textwidth}
         \includegraphics[width=6cm]{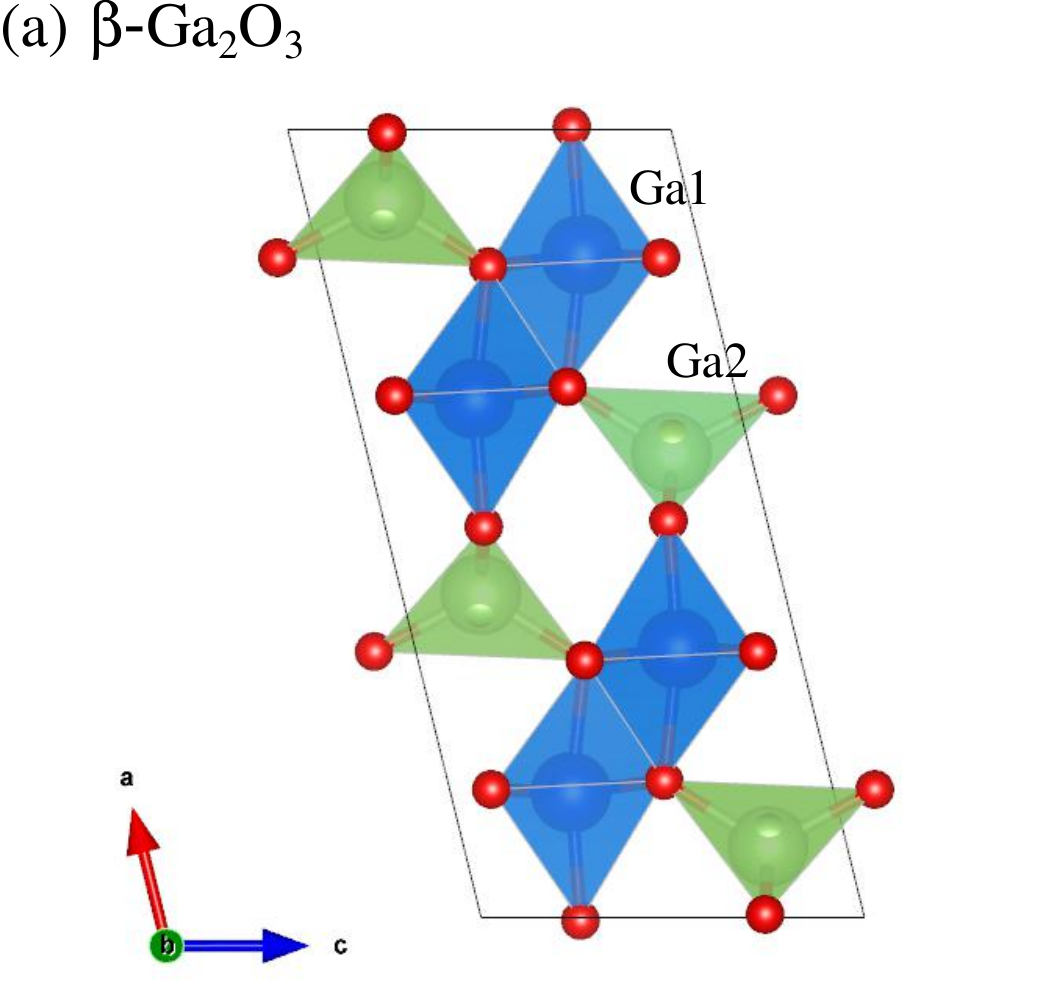}
         \label{fig:beta-unit}
     \end{subfigure}
     \begin{subfigure}{0.49\textwidth}
         \includegraphics[width=6cm]{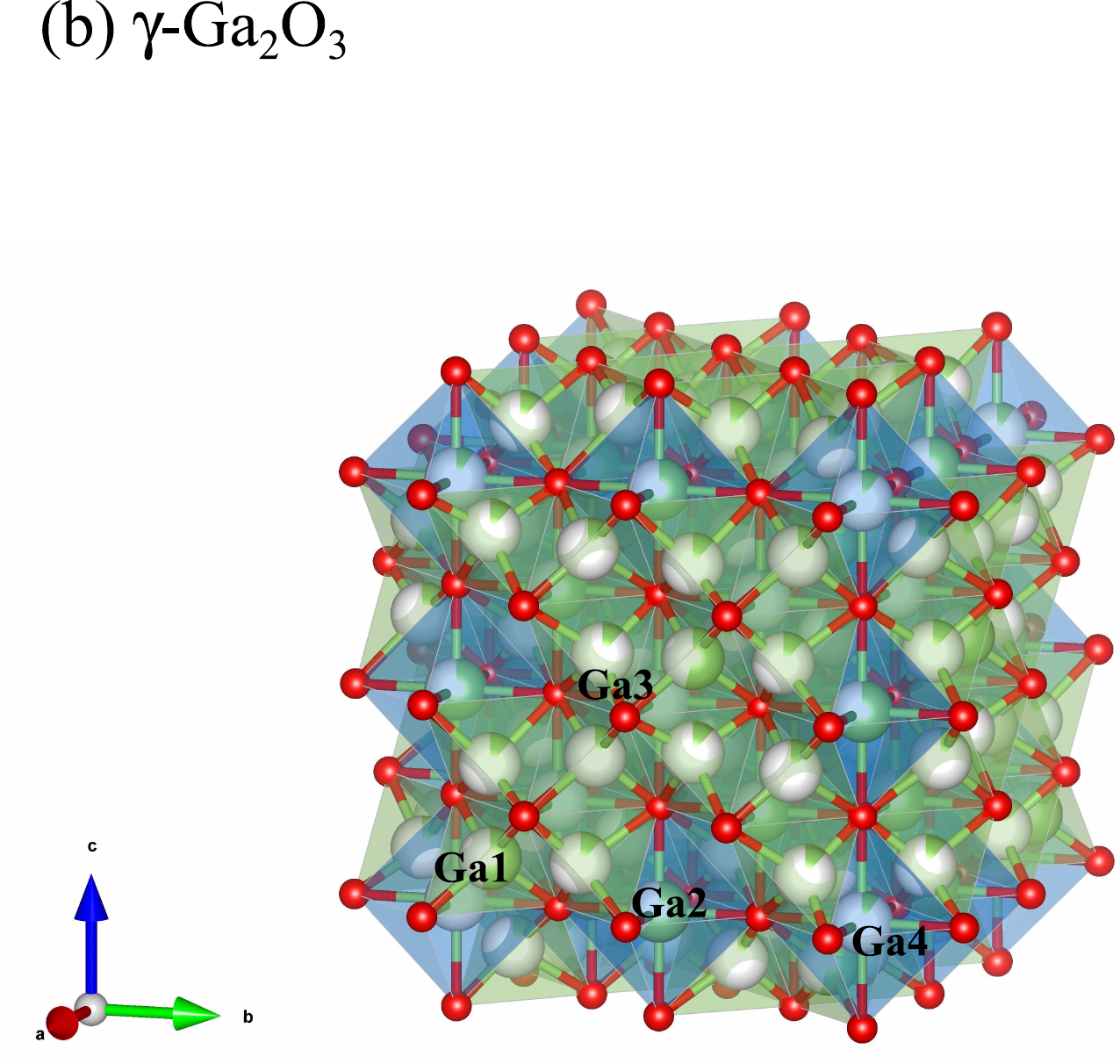}
         \label{fig:gamma-unit}
     \end{subfigure}
        \caption{Unit cell illustrations: (a) $\beta$-\ce{Ga2O3}, with red, blue, and green spheres representing O atoms, $\beta$-Ga$_\mathrm{Oct}$ at the octahedral center, and $\beta$-Ga$_\mathrm{Tetr}$ at the tetrahedral center, respectively. (b) $\gamma$-\ce{Ga2O3}, with green/white spheres visualizing e Ga atoms at possible sites with different occupancies. 
        }
        \label{fig:unit-cell}
\end{figure*}

$\beta$-\ce{Ga2O3} with space group $C2/m$ has a monoclinic crystal system, where $\angle \beta = 103.42 \degree$~\cite{kohn1957characterization}. 
The conventional cell of $\beta$-\ce{Ga2O3} contains 20 atoms. As shown in Fig.~\ref{fig:unit-cell}(a), there are two nonequivalent Ga sites ($\beta$-Ga$_\mathrm{Oct}$ bonded to the six O anions forming octahedral structures, and $\beta$-Ga$_\mathrm{Tetr}$ is bonded to four O ions forming tetrahedral structures), and consequently three nonequivalent O sites. 
Cubic $\gamma$-\ce{Ga2O3} (space group $Fd\overline{3}m$) crystal structure is illustrated in Fig.~\ref{fig:unit-cell}(b). 
O anions are expected to occupy the ideal spinel sites. 
In contrast, Ga cations partially occupy tetrahedral 8a ($\gamma$-Ga1), octahedral 16d sites ($\gamma$-Ga2), tetrahedral 48f ($\gamma$-Ga3), and octahedral 16c ($\gamma$-Ga4) sites with a refined tetrahedral to octahedral ratio of $1:1.35$~\cite{playford2013structures, playford2014characterization}. Because of this partial occupation, a $1\times1\times3$ supercell containing 160 atoms is required to account for all possible sites of the Ga sublattice. Our interface models use the most stable 3-site $\gamma$-\ce{Ga2O3} (IIIA) as reported by Ratcliff \textit{et al.}~\cite{ratcliff2022tackling}, since its bandgap matches the experimental value~\cite{new} best among structures they studied. In this model, Ga cations are at $\gamma$-Ga1, $\gamma$-Ga2, and $\gamma$-Ga3 sites with occupancies of 0.833, 0.833, and 0.028, respectively. Further details of the crystal structures of $\gamma$-\ce{Ga2O3} can be found in Ref.~\cite{ratcliff2022tackling}.

\begin{figure*}
     \centering
     \begin{subfigure}{0.49\textwidth}
         \includegraphics[width=8.6cm]{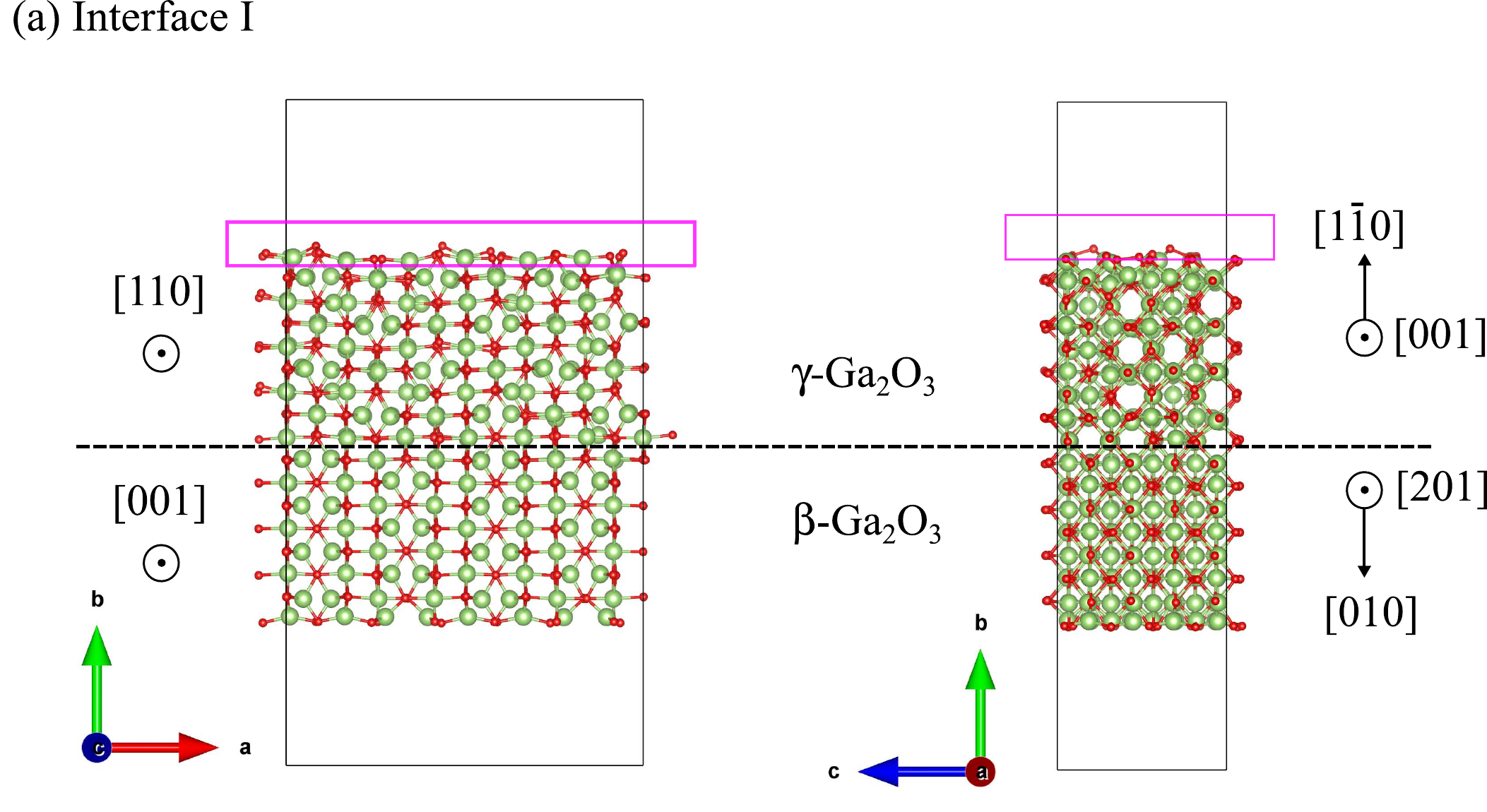}
         \label{fig:interface1}
     \end{subfigure}
     \begin{subfigure}{0.49\textwidth}
         \centering
         \includegraphics[width=8.6cm]{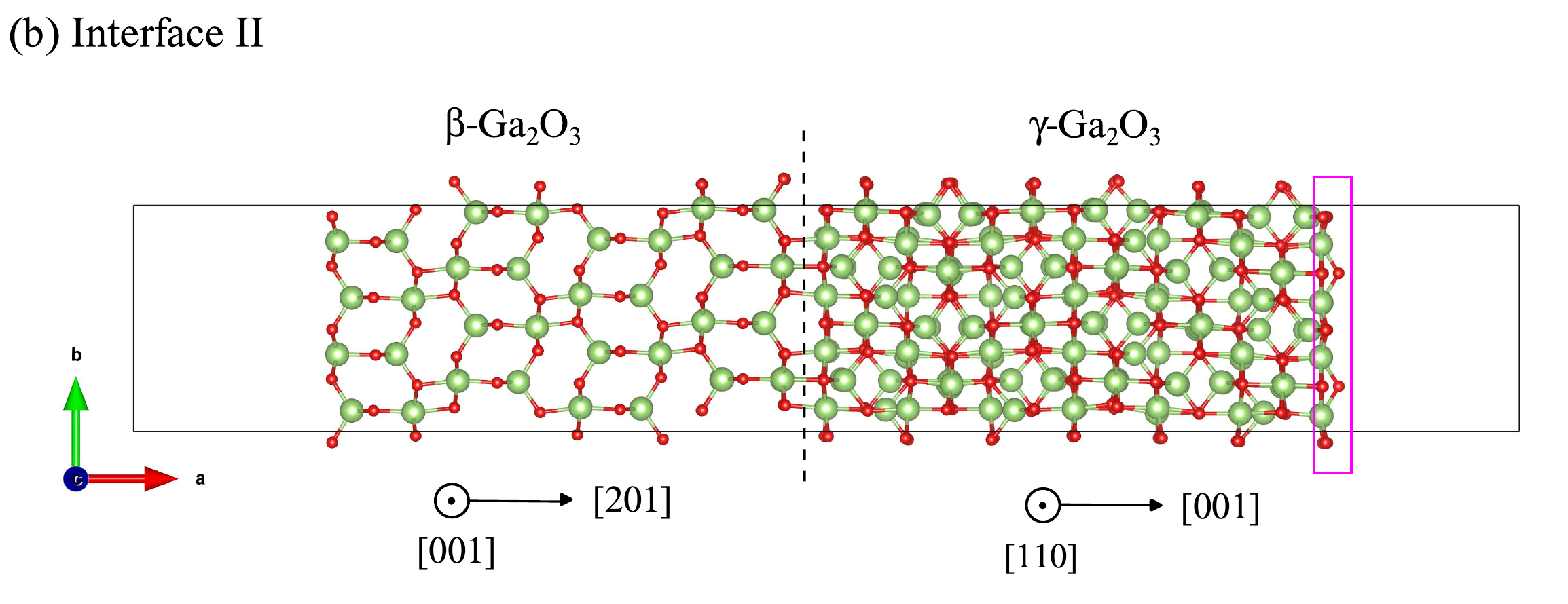}
         \label{fig:interface2}
     \end{subfigure}

        \caption{Interface structures studied in this work. The magenta boxes highlight the passivation layers.}
        \label{fig:interface}
\end{figure*}

Notably, several stacking orientations of the $\beta / \gamma$-\ce{Ga2O3} interface were reported in the literature, e.g., in~\cite{SFazarov2023universal}, using scanning transmission electron microscopy (STEM) and electron backscatter diffraction (EBSD) analyses, namely, $\gamma[100]\Vert\beta[201]$, $\gamma[110]\Vert\beta[132]$, and $\gamma[112]\Vert\beta[102]$.
Here, we focus on the $\gamma[100]\Vert\beta[201]$ stacking (referred in the following as Interface \Romannum{1}), see Fig.~\ref{fig:interface}(a). 
The $\gamma$-\ce{Ga2O3} structure is isotropic, thus $\gamma[100]$ is identical to $\gamma[001]$. 
Modeling the other two stacking orientations would require supercells containing more than 1000 atoms, which is beyond the state-of-the-art DFT calculation capacity and thus out of the scope of this work. 
In addition, we study the $\gamma[110]\Vert\beta[001]$ stacking orientation (Interface \Romannum{2}), see Fig.~\ref{fig:interface}(b), even though it was not experimentally detected yet, but was possible to compute. 
Interfaces \Romannum{1} and \Romannum{2} are built out of the same $\beta$ and $\gamma$ cells but joined in different crystallographic directions. 
A vacuum of 20~\r A is added to the interface structure to eliminate the influence of the other interfaces because of the periodic boundary conditions applied in all directions. 
An extra layer was added to the open surface of the $\gamma$-side to eliminate the polarization and nonphysical charge transfer. As a result, our models became nonstoichiometric but charge-neutral. 
320 atoms of the $\beta$-\ce{Ga2O3} cell and 356 atoms of the $\gamma$-\ce{Ga2O3} cell are used to construct Interface \Romannum{1}, while 344 atoms of the $\gamma$-\ce{Ga2O3} cell are used for Interface \Romannum{2}. 
Section Method provides more details on the surface termination. 

To create a coherent interface, it is necessary to strain the $\gamma$-cell to match the $\beta$-\ce{Ga2O3} lattice due to the difference in structures of the phases. The strain is applied to the $\gamma$-cell since the $\beta$-phase is more stable~\cite{mu2019ab}, while the $\gamma$-phase is transformed from $\beta$ in the experiments~\cite{huang2023atomic}. 
The $\beta$ and $\gamma$ cells adhere along the $a$-$c$ plane forming Interface \Romannum{1}, the $\gamma$-cell is thus compressed in $a$- and $c$-directions to match with the $\beta$ phase, while the $b$ lattice parameter is relaxed. In the following, we refer to this cell as Strained $\gamma$-cell-1.
Similarly, the $\gamma$-cell is compressed in $b$- and $c$-directions and relaxed along $a$-direction to form Interface \Romannum{2} (Strained $\gamma$-cell-2).

The GGA-PBE-optimized (Perdew-Burke-Ernzerhof version of generalized gradient approximation~\cite{perdew1996generalized}) lattice parameters and the corresponding bandgaps ($E_g$) of $\beta$-\ce{Ga2O3} unit cell, $\gamma$-\ce{Ga2O3} $1\times1\times3$ supercell, and cells constructing our interface models are listed in Table~\ref{tab:bandgap}. 
GGA-PBE and HSE (Heyd-Scuseria-Ernzerhof hybrid functional~\cite{hse2006ori}) can optimize the crystal to slightly different shapes and predict a considerably different volume. Thus, even though HSE-optimized lattice parameters or bandgap match better with the experimental ones, we fixed the lattice parameters to the GGA-PBE-optimized values for consistency. The slab and interface calculations were performed with GGA-PBE because these structures were too large to perform HSE calculations. 

The $E_\mathrm{g}$-HSE values listed in Table~\ref{tab:bandgap} are bandgaps calculated using HSE based on the GGA-PBE-optimized lattice, which are subsequently used for band offset calculations. This calculation approach tends to underestimate $E_\mathrm{g}$ for $\beta$ and $\gamma$-\ce{Ga2O3} compared to experimental and literature data. In contrast, bandgaps calculated using HSE-optimized structures (4.82 eV for $\beta$-\ce{Ga2O3} and 4.67 eV for $\gamma$-\ce{Ga2O3}) show better agreement with experimental results \cite{new}.

For analyzing strained $\gamma$ cells, the lattice parameter perpendicular to the matching surfaces is adjusted to alleviate the compressive strain experienced by the $\gamma$ phase in both interface models. As detailed in Table~\ref{tab:bandgap}, the relaxation along the \textbf{b}-axis in $\gamma$-cell-2 effectively reduces the strain to just --0.06\%, leading to a moderate bandgap narrowing. In contrast, $\gamma$-cell-1 maintains a substantial compressive strain of --3.17\%, which results in an increased bandgap as compared to the $\gamma$-cell in its equilibrium strain state. This difference in strain significantly influences the conduction band offset, highlighting the importance of strain management in optimizing electronic properties of the $\beta / \gamma$ interfaces.

\begin{table*}[bt]
\caption{Lattice parameters and bandgap of \ce{Ga2O3} polymorphs in accordance with the interface models in comparison with literature values. $\delta lnV$ represents fractional volume change of $\gamma$-cells. $E_\mathrm{g}$-HSE were obtained with HSE but used the GGA-PBE-optimized structures, while the literature values, $E_\mathrm{g}$-Other HSE, are the HSE-optimized result. }\label{tab:bandgap}
\begin{threeparttable}
\begin{tabular}{llllllll}
\hline
\hline
\thead{}    & \thead{$\beta$-\ce{Ga2O3}} & \thead{$\gamma$-\ce{Ga2O3}} & \thead{$\beta$-cell} & \thead{$\gamma$-cell} & \thead{Strained $\gamma$-cell-1} & \thead{Strained $\gamma$-cell-2}  & \thead{$\gamma$-cell-GPA}\\
\hline
\textit{a} (\AA) &12.47  &8.40   & 24.23     &25.21   &24.23     &24.48     &24.23\\
\textit{b} (\AA) &3.09   &8.40   & 12.35    &11.88   &12.09     &12.35     &11.89\\
\textit{c} (\AA) &5.88   &25.21  & 11.77   &11.88   &11.77     &11.77     &11.77\\
$\delta ln(V)$ (\%) &      &       &         & 0       & --3.17         & -0.06       &  --4.70     \\
$E_\mathrm{g}$-GGA (eV)    & 1.97     & 1.74        &     &   &2.07     &1.58       &2.07\\                               
$E_\mathrm{g}$-HSE (eV)    & 4.38     & 4.13        &     &   &4.31    &4.03        &4.52\\                               
$E_\mathrm{g}$-Expt. (eV)  & 4.7$\sim$4.9 \textcircled{1}   &5.0 \textcircled{2} & & & & & \\
$E_\mathrm{g}$-Other HSE (eV) & 4.92 \textcircled{3}, 4.8 \textcircled{4}     & 4.69 \textcircled{5} & & & & &  \\  
\hline  
\hline  
\end{tabular}

\begin{tablenotes}
\item \textcircled{1} Refs.~\cite{matsuzaki2006growth, stepanov2016gallium, kohn1957characterization, he2006electronic, orita2000deep}; 
\item \textcircled{2} Ref.~\cite{oshima2012epitaxial};
\item \textcircled{3} Ref.~\cite{gake2019first};
\item \textcircled{4} Ref.~\cite{deak2017choosing};
\item \textcircled{5} Ref.~\cite{ratcliff2022tackling}
\end{tablenotes}
\end{threeparttable}
\end{table*}

\subsection{Band offsets}\label{bandsoff-calc}
The interface band offsets were firstly determined through the potential lineup method~\cite{baldereschi1988band, colombo1991valence, peressi1998band}. 
The potential lineup method requires an interface calculation and two bulk calculations for two interface components.
The VBM offset (VBO) for $\beta / \gamma$ interface can then be determined via the following equation:
\begin{equation} \label{eq:1}
\mathrm{VBO}_\mathrm{(\beta / \gamma)} = (\epsilon_\mathrm{V}^\mathrm{\beta}-\overline{V}^\mathrm{\beta})-(\epsilon_\mathrm{V}^\mathrm{\gamma}-\overline{V}^\mathrm{\gamma})+(\overline{V}^\mathrm{\beta}-\overline{V}^\mathrm{\gamma}),
\end{equation}
where $\epsilon_\mathrm{V}^i$ refer to the VBM obtained from the bulk calculations, $\overline{V}^i$ is the macroscopic averaged electrostatic potential of two components, obtained by converging the planar averaged potential with a moving average with $i = \beta$ or $\gamma$ respectively. As such $\overline{V}^\mathrm{\beta}-\overline{V}^\mathrm{\gamma}$ denotes the interface lineup determined in the interface heterostructure. 
The corresponding CBM offset (CBO) can then be calculated via:
\begin{equation} \label{eq:2}
\mathrm{CBO}_\mathrm{(\beta / \gamma)} = (E_\mathrm{g}^\mathrm{\beta}-E_\mathrm{g}^\mathrm{\gamma})+ \mathrm{VBO}_\mathrm{(\beta/\gamma)}
\end{equation}
where $E_\mathrm{g}^i$ is the bandgap of $\beta$ and $\gamma$-\ce{Ga2O3}, respectively, as listed in Table~\ref{tab:bandgap} which is referred as $E_\mathrm{g}$-HSE.
Considering expensive computational costs, the band structure components $\epsilon_\mathrm{V}^i$ and lineup \(\overline{V}^\mathrm{\beta}-\overline{V}^\mathrm{\gamma}\) were calculated using GGA~\cite{kresse1994norm, kresse1999ultrasoft, perdew1996generalized}, while the bandgaps $E_\mathrm{g}^i$ were calculated with the hybrid functional~\cite{flores2018accuracy}.
According to the previous studies~\cite{lyu2023band, alkauskas2008band, weston2018accurate, steiner2014band, shaltaf2008band}, the use of GGA can yield almost the same interface lineup as that in the hybrid functional calculations.

\begin{figure*}
     \centering
     \begin{subfigure}{0.49\textwidth}
         \centering
         \includegraphics[width=8cm]{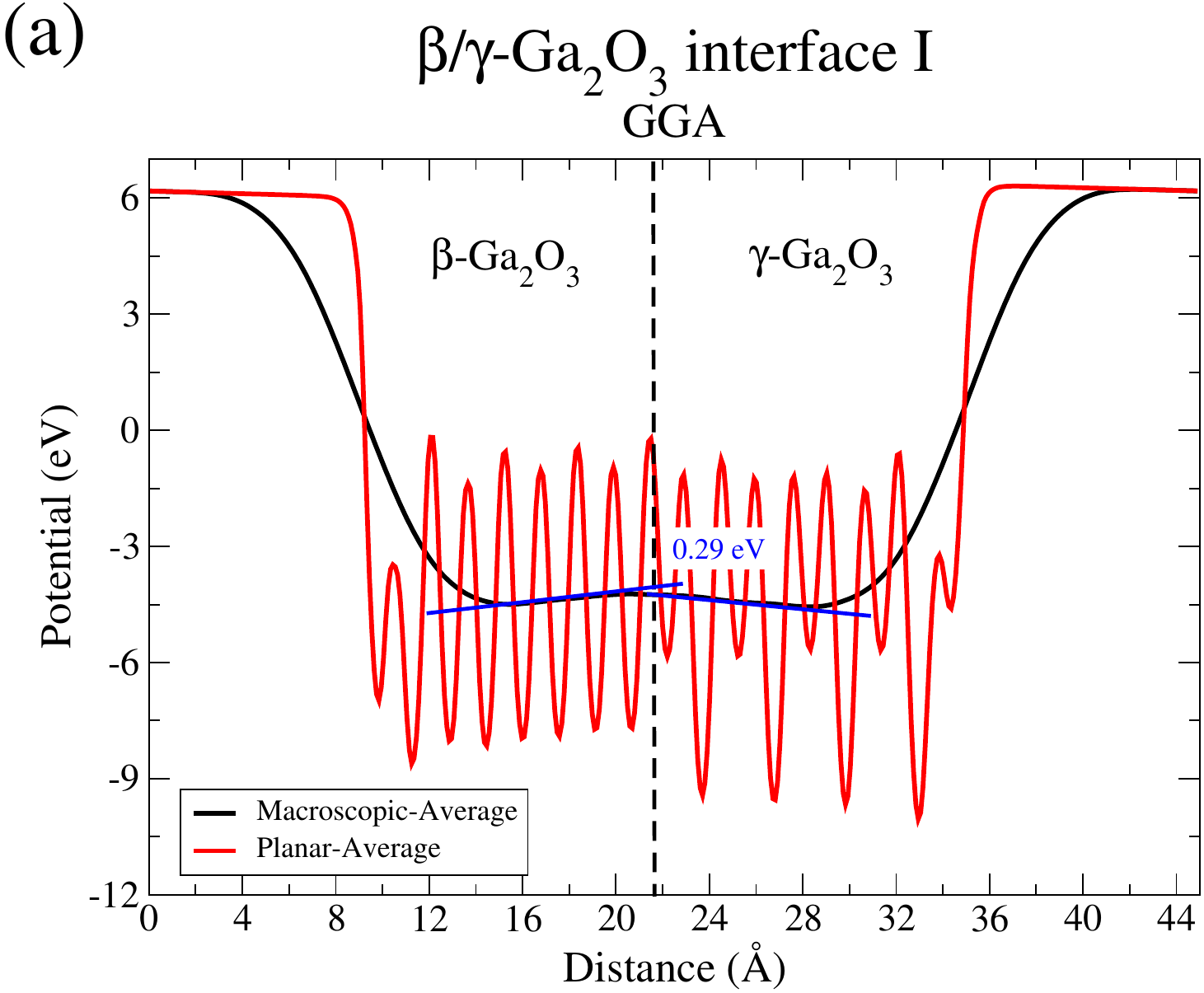}
         \label{fig:inter-potential-a}
     \end{subfigure}
     \begin{subfigure}{0.49\textwidth}
         \centering
         \includegraphics[width=8cm]{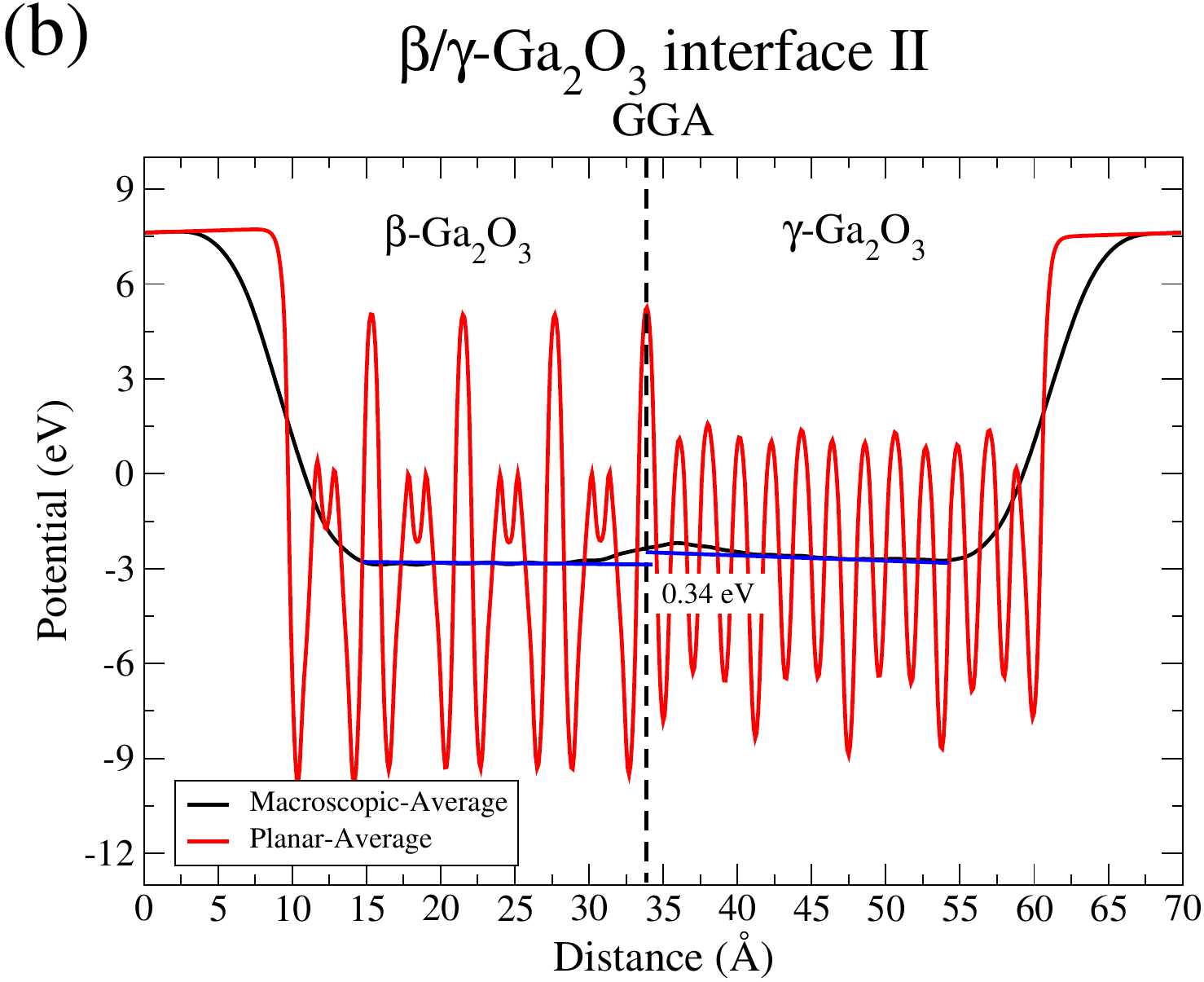}
         \label{fig:inter-potential-b}
     \end{subfigure} 
        \caption{Macroscopic average potentials (black curves) and planar average potentials (red curves) of $\beta / \gamma$-\ce{Ga2O3} interfaces. Blue lines highlight the extrapolations at the nominal interface position. The lineup values are also marked there.}
        \label{fig:inter-potential}
\end{figure*}

The planar potential profile of the interfaces is extracted by using the data post-processing code VASPKIT~\cite{VASPKIT} as shown in Fig.~\ref{fig:inter-potential}. The macroscopic average potential and planar average potential represent the electrostatic potential averaged along the \textit{b}-axis for Interface \Romannum{1} and along the \textit{a}-axis for Interface \Romannum{2}.
In our models, the asymmetric slabs give rise to the finite electric fields across the interfaces 
and vacuum because of the periodic boundary conditions~\cite{bengtsson1999dipole}.
We employ the extrapolation scheme developed by Foster \textit{et al.}~\cite{foster2014band} to eliminate the effect of the built-in electric fields.
This extrapolation scheme has been successfully applied to band offsets in other $\beta$-\ce{Ga2O3} interfaces~\cite{lyu2023band, lyu2020band}. 
The difference between the two extrapolations at the nominal interface position is regarded as the interface lineup, \(\overline{V}^\mathrm{\beta}-\overline{V}^\mathrm{\gamma}\). 
Here, we define the nominal interface position as the midpoint between the $\beta$ and $\gamma$ surfaces.
Using this procedure, we determined the lineups of 0.29 eV and --0.34 eV for Interfaces \Romannum{1} and \Romannum{2} respectively. 
With these lineup values, we calculated the VBO and CBO using Eqs.~(\ref{eq:1}) and (\ref{eq:2}). 
The results are presented in Table~\ref{tab:bandoffset}. 

In addition to the interface calculation, the band offset can be extracted from the slab calculations by vacuum level alignment~\cite{mu2020orientation}. 
This method regards the potential of vacuum as reference ($V_\mathrm{vac} = 0$). It combines the positions of band edges obtained in the bulk calculations ($\epsilon_\mathrm{V}$ or $\epsilon_\mathrm{C}$) with the averaged potentials of slabs ($V_\mathrm{av}$). 
In this way the VBM position the absolute energy scale is then defined as:
\begin{equation} \label{eq:3}
\mathrm{VBM} =\epsilon_\mathrm{V} + V_\mathrm{av}. 
\end{equation}
It is worth noting that, in both of the above methods, atomic relaxation and strain play significant roles in determining $V_\mathrm{av}$ and the band edge positions relative to it. It is important to always use bulk reference cells that are strained consistently with the strain of the corresponding interface or the slab model. Here, we applied strain to $\gamma$-cells and the atomic relaxation is performed at GGA level. 


\begin{figure}
     \centering
     \begin{subfigure}{0.49\textwidth}
         \centering
         \includegraphics[width=8cm]{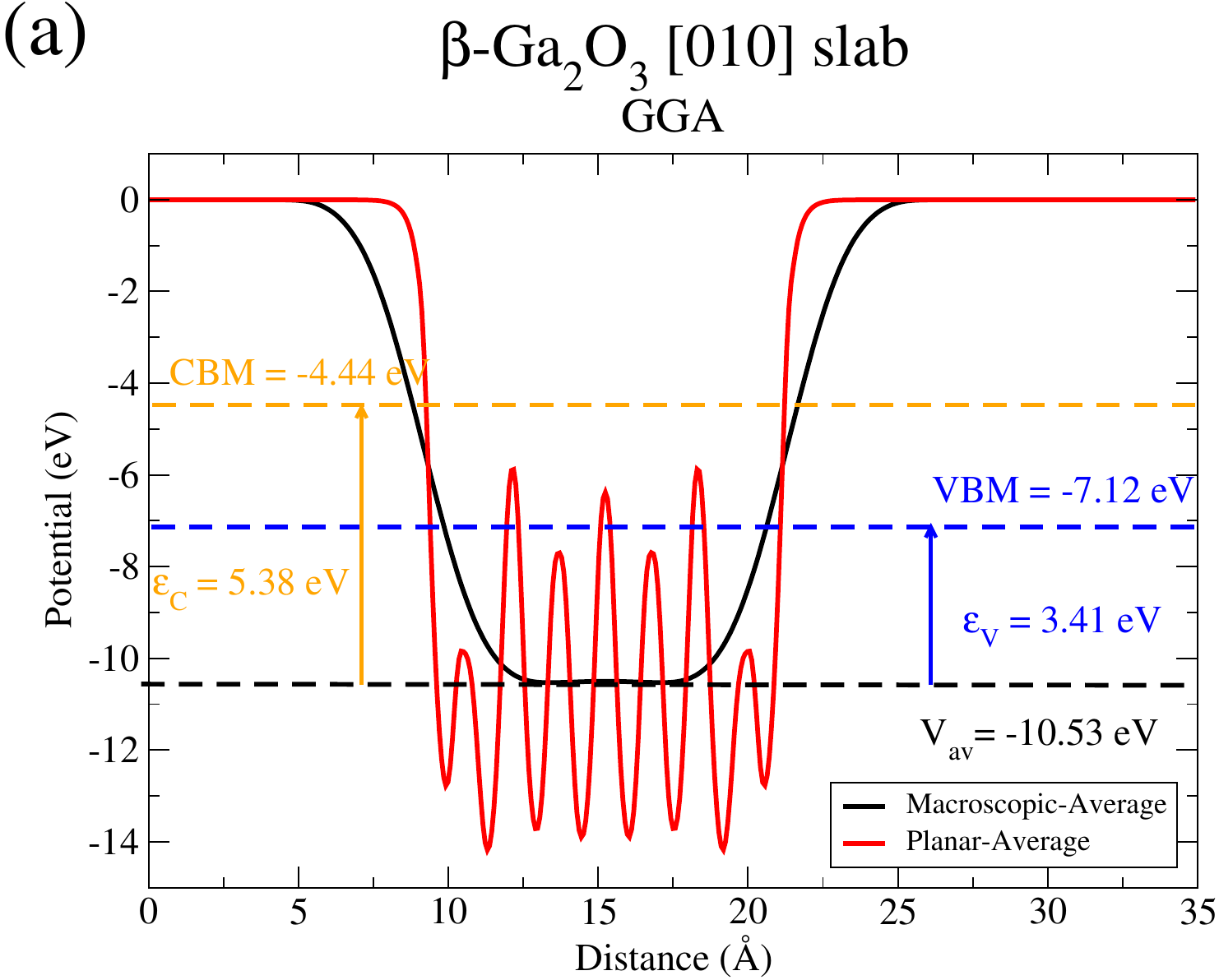}
         \label{fig:potential-a}
     \end{subfigure}
     \begin{subfigure}{0.49\textwidth}
         \centering
         \includegraphics[width=8cm]{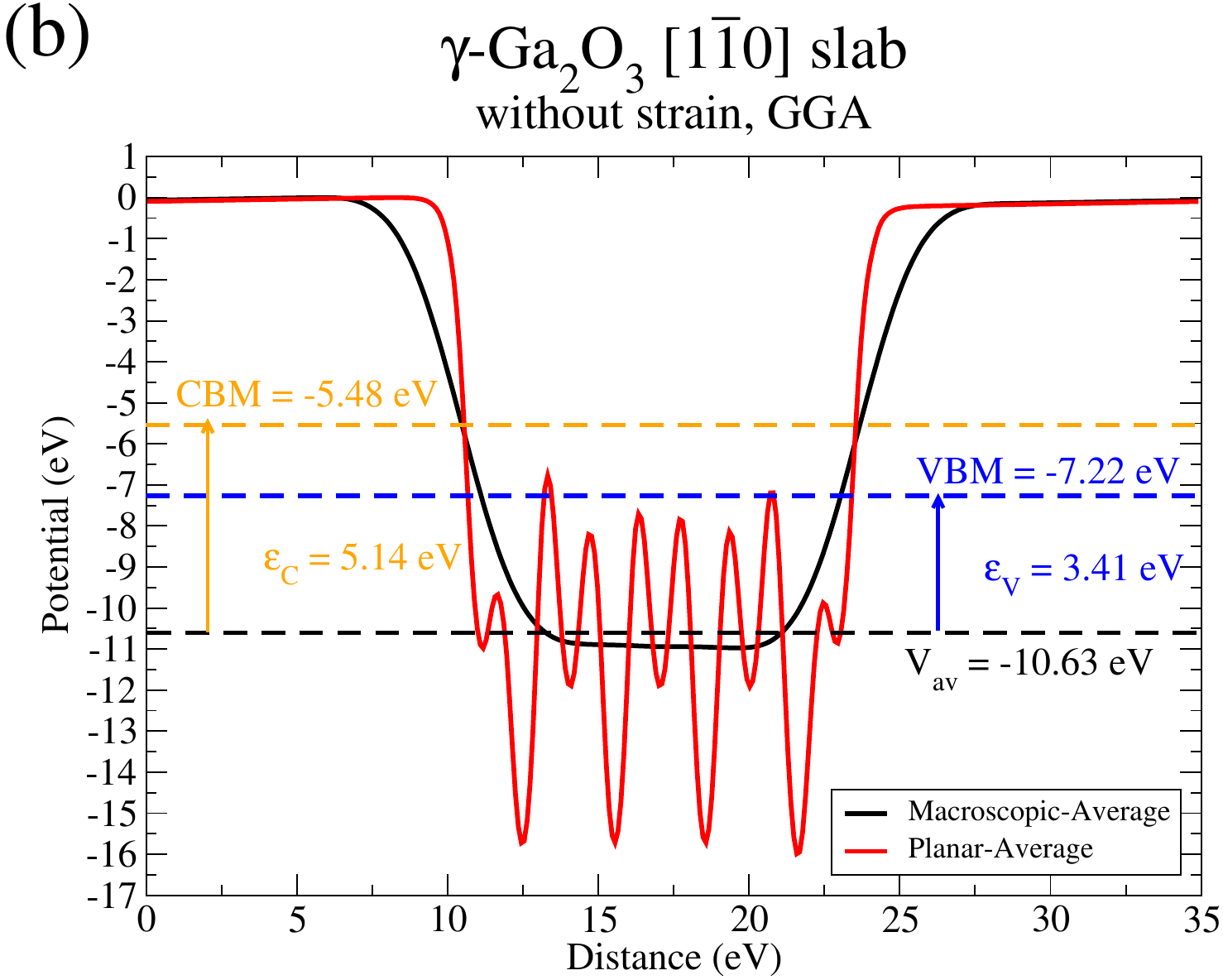}
         \label{fig:potential-b}
     \end{subfigure}
        \caption{Determination of the band edge positions with respect to $V_\mathrm{av}$, for the example of the $\beta[010]$ and $\gamma[1\overline{1}0]$ slabs which constructs Interface \Romannum{1}. Here $\gamma$-\ce{Ga2O3} slab is in an equilibrium state. $\epsilon_\mathrm{V}$ and $\epsilon_\mathrm{C}$ are obtained from bulk calculations.}
        \label{fig:potential}
\end{figure}

By employing the vacuum level as a reference, we determined the VBM positions of the $\beta$ and $\gamma$ phases in the interface using Eq.~(\ref{eq:3}), as illustrated in Fig.~\ref{fig:potential}, thereby determining the VBO.
Instead of using the same scheme, we compute the CBO using Eq.~(\ref{eq:2}) to mitigate possible underestimation issues associated with GGA. 
The strain-applied cases are also studied to ensure the comparability of these two methods. 

\begin{figure}
    \centering
     \begin{subfigure}{0.49\textwidth}
         \centering
         \includegraphics[width=8cm]{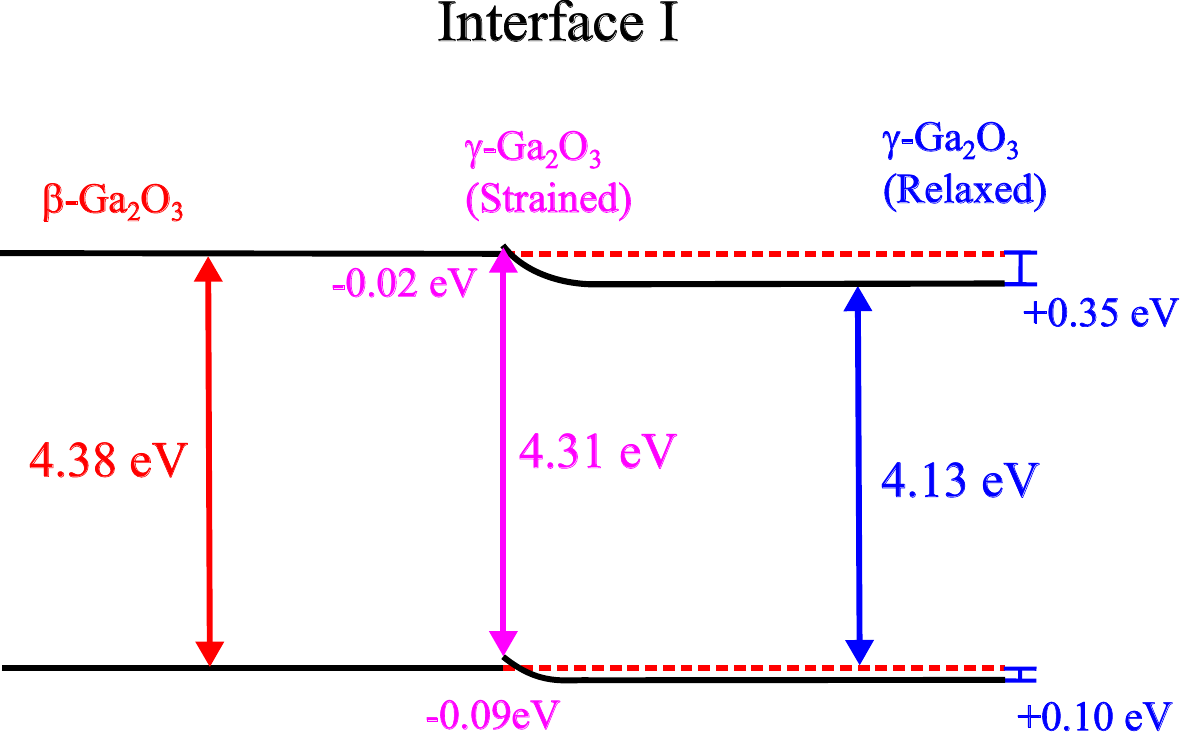}\\
     \end{subfigure}
     \begin{subfigure}{0.49\textwidth}
         \centering
         \includegraphics[width=8cm]{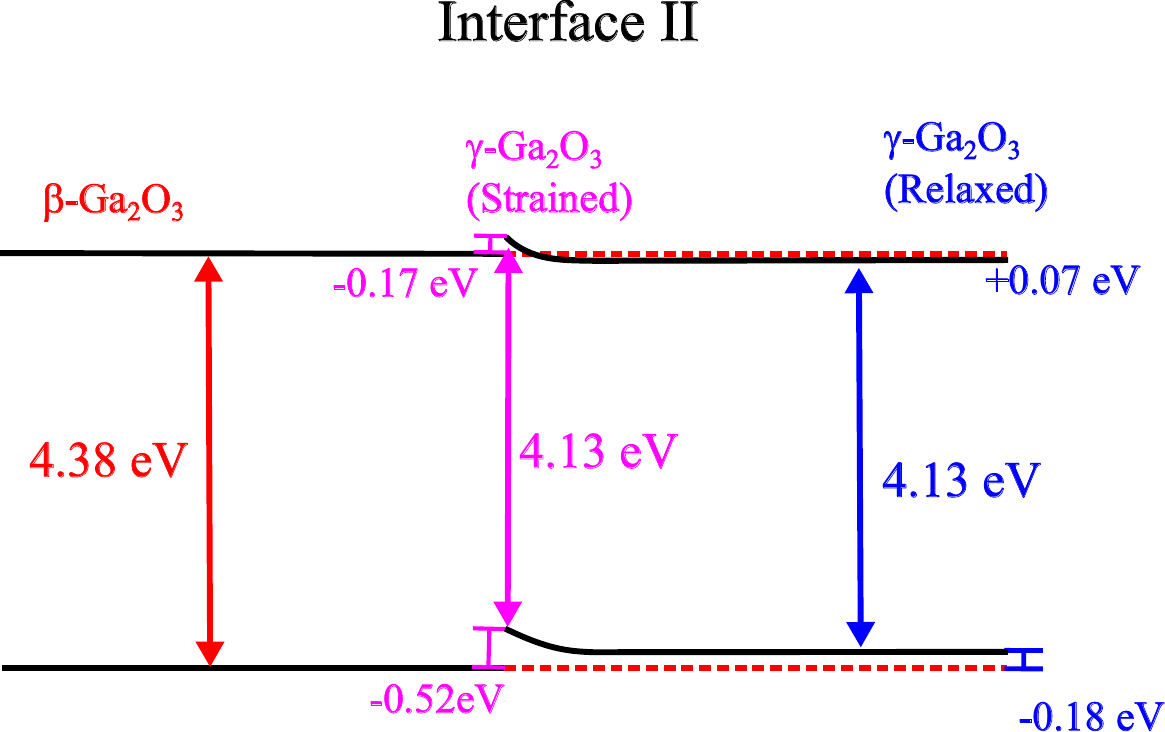}
     \end{subfigure}    
    \caption{Band offset diagram of the interfaces between $\beta$- and $\gamma$-\ce{Ga2O3}. 
    }
    \label{fig:bandoffset}
\end{figure}

As seen from Table~\ref{tab:bandoffset}, the vacuum alignment method and the potential lineup method yield perfectly consistent results. 
For interface \Romannum{1}, when compressive strain is applied to the $\gamma$ component, both CBO and VBO are comparable and negligible. However, in the absence of strain, the CBO increases significantly.
For interface \Romannum{2}, VBO becomes larger than CBO when strain is applied to the $\gamma$ component. In the absence of strain, CBM of $\gamma$ shifts downward while CBO moves in the opposite direction; despite this shift, CBO remains smaller than that with strain. This behavior illustrates the impact of strain on the electronic properties at the interfaces.
The overall band offset diagram of $\beta / \gamma$-\ce{Ga2O3} interface is shown in Fig.~\ref{fig:bandoffset}

\begin{table*}[bt]
   \centering
\caption{The band offsets of $\beta / \gamma$-\ce{Ga2O3} interfaces calculated with different methods.} \label{tab:bandoffset}
\begin{threeparttable}
\begin{tabular}{lccrr}
\hline
\hline
\thead{}  & \multicolumn{2}{c}{\textbf{Interface \Romannum{1}}}        &\multicolumn{2}{c}{\textbf{Interface \Romannum{2}}}\\
\hline
\thead{Method} & \thead{CBO} & \thead{VBO}  & \thead{CBO} & \thead{VBO} \\
\hline
Potential lineup                  &0.00  & --0.08      & --0.17 & --0.53    \\
Vacuum alignment (with strain)     & --0.02  & --0.09      & --0.17 & -0.52  \\
Vacuum alignment (without strain)  & 0.35  & 0.10      &0.07 & --0.18  \\
\hline
\hline
\end{tabular}
\end{threeparttable}
\end{table*}

\section{Discussion}\label{sec:level4}

Our calculation predicts that the band offsets between $\beta$-\ce{Ga2O3} and $\gamma$-\ce{Ga2O3} are orientation-dependent, which has previously been observed at the interface between (Al$_x$Ga$_{1-x}$)$_2$O$_3$ and \ce{Ga2O3}~\cite{mu2020orientation}.
However, more electronic structure calculations of other stacking-orientated interface structures observed by previous work \cite{SFazarov2023universal} should be done to come to a conclusive statement. Meanwhile, it is interesting to highlight the effect of strain on the band offsets. The lateral compressive strain applied to the $\gamma$ phase can reduce its band gap, enhancing the offset (interface \Romannum{2}) or can increase the bandgap, reducing the offset (interface \Romannum{1}), and this effect depends on the slab orientation. Assuming $n$-type conductivity, the electron transport properties in \ce{Ga2O3} are governed by CBO. This means electrons at the $\beta / \gamma$-\ce{Ga2O3} interface tend to accumulate at the side with a lower CBM, suggesting the possibility of inducing two-dimensional electron gas (2DEG) by the strain on the $\gamma$-side of the interface.

\begin{figure}[h]
    \includegraphics[width=8cm]{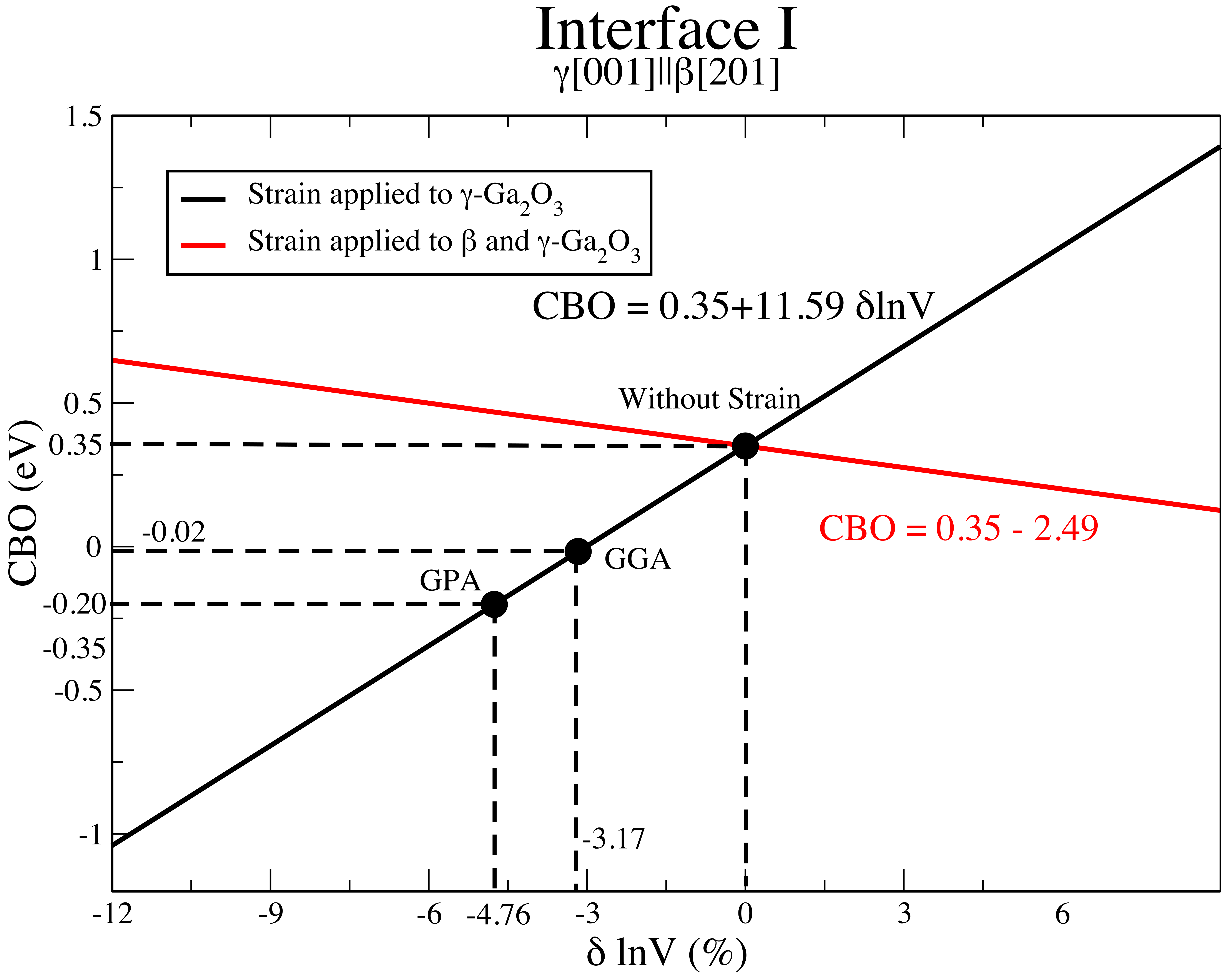}
    \caption{CBO of strained Interface \Romannum{1}. The black line represents the case when $\beta$-\ce{Ga2O3} is in equilibrium state and $\gamma$ in strain state, while the Red line represents the case when the same strain is applied to both $\beta$ and $\gamma$. Black dots highlight our calculated CBO with and without strain, and the extrapolated CBO at GPA observed strain state.}
    \label{strain}
\end{figure}

As a matter of fact, the lateral compressive strain results in a volume compression of $\gamma$-\ce{Ga2O3} even though the perpendicular axis is relaxed. The volume compression varies depending on the interface. For Interface \Romannum{1}, the $\gamma$-cell decreases by 3.17\% in volume, corresponding with $\Delta \mathrm{CBO} = 0.37$~eV. For interface \Romannum{2}, the volume of $\gamma$-cell decreases by only 0.06\%, resulting in $\Delta \mathrm{CBO} = 0.24$~eV. This indicates that the magnitude of the strain effect is also orientation-dependent. We quantify the strain effect by determining the value of the absolute deformation potential $a_c$ \cite{van1989absolute, mu2020orientation} for the CBM in $\gamma$-\ce{Ga2O3}: for [1$\overline{1}$0]-oriented slab $a_c = -11.59$~eV. Then, the relationship between the strain and CBO can be expressed as:
\begin{equation} \label{eq:4}
    \mathrm{CBO} = \mathrm{CBO}^{eq}-a_c \delta \ln{(V)}\
\end{equation}
where \(\delta \ln{(V)} = \delta V / V\) is the fractional volume change. If the same \(\delta \ln{(V)}\) is applied to $\beta$-\ce{Ga2O3}, CBO is then expressed as:
\begin{equation}\label{eq:5}
 \mathrm{CBO} = \mathrm{CBO}^{eq}-\Delta a_c \delta \ln{(V)}    
\end{equation} 
where \(\Delta a_c = a_c^\beta - a_c^\gamma\). Mu \textit{et al.}~\cite{mu2020orientation} have reported a value of --9.1 eV for the absolute deformation potential for a [010]-oriented $\beta$ slab. Combining literature and our data, we plot the impact of strain on CBO at interface \Romannum{1} (which is composed of [1$\overline{1}$0]-oriented $\gamma$ slab and [010]-oriented $\beta$ slab) in Fig.~\ref{strain}. Thus, keeping $\beta$-\ce{Ga2O3} in an equilibrium state and applying strain to $\gamma$ can significantly change the CBO. 
When the volume of $\gamma$ expands, CBO increases, enhancing chances for the 2DEG formation in $\gamma$-\ce{Ga2O3}. When the volume compresses, CBO reduces until the compression reaches 6.04 \%, where CBO becomes larger in the opposite direction, opening an opportunity for the 2DEG  formation in $\beta$-\ce{Ga2O3}. However, high strain magnitudes can also provoke phase instability, making too high strain adjustments impractical. 
Otherwise, applying the same strain to $\beta$ and $\gamma$, the effect becomes different since $\beta$ phase is less sensitive to strain than that of $\gamma$. The CBO increases slightly when $\beta$ and $\gamma$ are compressed and decreases slightly upon the expansion. In this case, the strain effect is minimal. 

\begin{figure*}
        \centering
    \includegraphics[width=16cm]{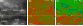}
    \caption{(a) High-resolution TEM image showing the interfacial area of the high-dose implanted (1×10$^{16}$ Ni/cm$^2$) sample, (b)-(c) the corresponding GPA $\varepsilon_{xx}$ (in-plane) and $\varepsilon_{yy}$ (out-of-plane) strain maps. The yellow arrow annotates the $\beta / \gamma$ interface.}
    \label{fig:GPA}
\end{figure*}

To unravel the real strain state of the $\beta / \gamma$-\ce{Ga2O3} interface, nanoscale strain analysis was conducted by applying geometric phase analysis (GPA) on high-resolution TEM images in Fig.~\ref{fig:GPA}. To start with, Fig.~\ref{fig:GPA}(a) shows a high-resolution TEM image of the interfacial region similar to that reported by Azarov \textit{et. al}~\cite{SFazarov2023universal}. To take this image, the sample was oriented along the [201] zone axis of the $\beta$-phase and the [001] zone axis of the $\gamma$-phase (idential with DFT modeled Interface \Romannum{1}). 
The orientation relationship between the two phases is described by $(20\overline{4})\beta\Vert(4\overline{4}0)\gamma$ (in-plane matching, planes vertical to the interface)
and $(020)\beta\Vert(440)\gamma$ (out-of-plane, planes parallel to the interface).

The relative deviations of the interplanar spacings \textit{\textbf{d}} of the $\gamma$-\ce{Ga2O3} film with respect to the $\beta$-\ce{Ga2O3} substrate (reference area) were found to be $\varepsilon_{xx} = 0$ and $\varepsilon_{yy} = -0.037$. Comparing to the nominal in-plane ($\varepsilon_{xx}$) and out-of-plane ($\varepsilon_{yy}$) lattice differences, we conclude that the $\beta$-\ce{Ga2O3} phase is in a equilibrium state, and the $\gamma$-\ce{Ga2O3} phase exhibits $-3.9\%$ in-plane compressive strain and $+0.1\%$ out-of-plane tensile strain.

The GPA results illustrated that, for the Interface \Romannum{1}, $\gamma$-\ce{Ga2O3} is compressively strained at the interface where its formation is initiated (in \textit{a}- and \textit{c}-axes) and expanded slightly in \textit{b}-axis (see $\gamma$-cell-GPA in Table~\ref{tab:bandgap}). The volume of $\gamma$-cell-GPA is compressed by 4.71\% compared to the equilibrium $\gamma$-cell. As it is modeled, the bandgap in such a cell increases up to 4.52 eV due to this volume compression compared with the equilibrium $\gamma$-\ce{Ga2O3}, which exhibits a bandgap of 4.13 eV. 
This may be interpreted as a consistent theory-experiment trend, even though the calculation for the strained $\gamma$-cell-1 overestimated the out-of-plane tensile strain, leading to a slightly larger $b$ lattice parameter than that in the GPA observation in Fig.\ref{fig:GPA}. 
This overestimation is likely because of the exchange-correlation functional limitations, so GGA-PBE overestimates the crystal volume. 
Additionally, extended defects such as dislocations and/or defect clusters potentially occurring in the sample may affect the strain, which was not considered in the computational model. At this end, we concluded that the GPA results support our theoretical predictions that $\gamma$ phase exhibits out-of-plane tensile strain at the $\beta / \gamma$ interface. 

Keeping in mind this consistent result, we have to admit that accurate strain state modeling at the interfaces based on GPA results or HSE approaches involves significant challenges. The strain state evolves during the slab relaxation processes, typically performed using GGA-PBE due to the high computational costs of more precise methods. Nevertheless, the slab relaxation is crucial when determining the band offset through the vacuum alignment method. As Mu \textit{et al.}~\cite{mu2020orientation} have shown, atomic relaxation can significantly alter the average electrostatic potential ($V_\mathrm{av}$), leading to shifts in the band edge positions, as illustrated in Fig.~\ref{fig:potential}.
Despite these complexities, our study represents perhaps the first successful attempt to evaluate band offsets in $\beta / \gamma$ polymorphic heterostructure directly by connecting theory and experiment. 

\section{Conclusions}\label{sec:level5}
In this work, we have studied band offsets of $\gamma[001]\Vert\beta[201]$ and $\gamma[110]\Vert\beta[001]$ \ce{Ga2O3} interface models using first-principles calculations validated by atomistic interface structure observed experimentally. As a result, we obtained realistic estimates for the band offsets, emphasizing the role of strain and crystallographic orientation. Specifically, we conclude that the band offsets between the $\beta$ and $\gamma$ phases are predicted to be relatively small, not exceeding a few hundred meV; however, they are highly sensitive to the strain state at the interface. This suggests that even minor variations in strain can significantly impact the electronic properties of the interface.  Similarly, the offset predictions strongly depend on the crystallographic orientation and even the computational model used. This variability underlines the importance of comparing the computations with experimental data. At this end, we conclude that even though the formation of a 2DEG at the $\beta / \gamma$ interface is theoretically possible, the gradual strain relaxation—if it occurs as a function of the distance from the interface—poses a significant challenge, as it may shift the 2DEG location or even reduce the overall probability of its formation.

\section{Technical Section}\label{method}
\subsection{DFT Calculation details}
Our DFT calculations are performed using the Vienna \textit{ab initio} simulation package (\textsc{VASP}, v.6.3.2)~\cite{kresse1996efficient} with the implemented projector augmented wave (PAW) potentials~\cite{kresse1994norm,kresse1999ultrasoft} and generalized gradient approximation (GGA): Perdew-Burke-Ernzerhof (PBE) exchange–correlation functional~\cite{perdew1996generalized}. 
The $3d$ electrons of Ga are treated as valence, as it has been confirmed by previous study~\cite{mu2019ab} that this setting can reproduce the ordering of phase stability correctly. 
To capture the electronic properties accurately and obtain the bandgap close to the experimental value, we employ the Heyd-Scuseria-Ernzerhof hybrid functional (HSE)~\cite{hse2006ori} incorporating a modified fraction of Hartree-Fock exchange ($\alpha=0.32$)~\cite{lyons2019electronic}.
The calculations are carried out with a plane-wave cutoff energy of 400 eV. For the integration in the Brillouin zone, the Gaussian smearing method~\cite{jorgensen2021effectiveness} is applied with a width of 0.05 eV. The electronic system is relaxed to reach a stable state, keeping the cell shape and volume constant. During this relaxation ions are allowed to move only until the forces between them are reduced to below 0.01 eV/\r A. The Brillouin zone is sampled with a $\Gamma$-centered mesh and the \textbf{k}-spacing value to generate \textbf{k}-mesh is 0.04 \r A$^{-1}$ which is equivalent to a dense $2\times8\times4$ for $\beta$-\ce{Ga2O3} unit cell and $3\times3\times1$ for $\gamma$-\ce{Ga2O3} $1\times1\times3$ supercell. For large slab and interface supercells ($> 300$ atoms), the Brillouin zone is sampled by the $\Gamma$-point~\cite{lyu2023band}. $\beta$ and $\gamma$-\ce{Ga2O3} exhibit indirect bandgaps, with the valence band maxima (VBMs) at the $\Gamma$-point. Consequently, the CBO is determined using the smaller unit cell and the method elaborated below (refer to Eq.~(\ref{eq:2})), which serves to address the limitation of using the $\Gamma$ point only. 

\subsection{Surface termination}
In this work, the surface termination scheme plays a crucial role in determining the band offsets, particularly for $\gamma$-\ce{Ga2O3}.
We note that the surface ions reorganize during the GGA relaxation of the slab calculations.
Vacuum added to the system to introduce the surface breaks the Ga-O bonds, distorting the symmetry of tetrahedral and octahedral positions of Ga cations. Ga in the center of tetrahedra becomes bonded to another Ga, forming unreasonable Ga-Ga bonds while octahedral Ga relaxes to the tetrahedral sites without forming extra bonds.
Thus it is essential to terminate the surface in such a way that only Ga$_\mathrm{Oct}$ are exposed to vacuum at the open surface.
However, it is impossible to have Ga$_\mathrm{Oct}$ at both the top and the bottom of the stoichiometric $\gamma$-slab.
Hence, an extra layer of Ga$_\mathrm{Oct}$ and O is added for termination of the surface facing vacuum (from here on we call this octa-Ga termination). To verify our approach, we studied the octa-Ga surface termination on the example of the [001]-oriented $\gamma$ slab of the 2-site (IIA) $\gamma$-\ce{Ga2O3} model.

\begin{figure*}
     \centering
     \begin{subfigure}{0.49\textwidth}
         \centering
         \includegraphics[width=8cm]{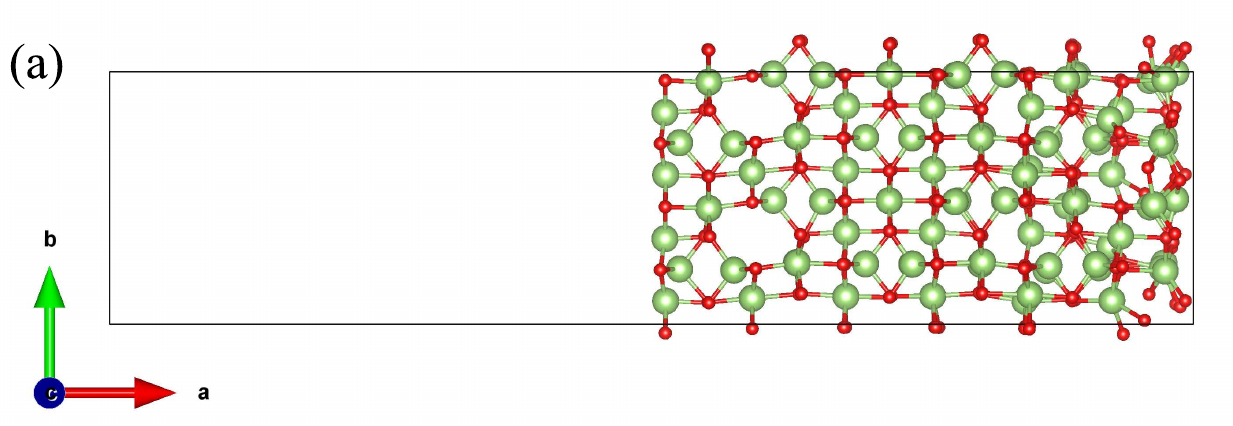}
         \label{fig:terminate-a}
     \end{subfigure}
     \begin{subfigure}{0.49\textwidth}
         \centering
         \includegraphics[width=8cm]{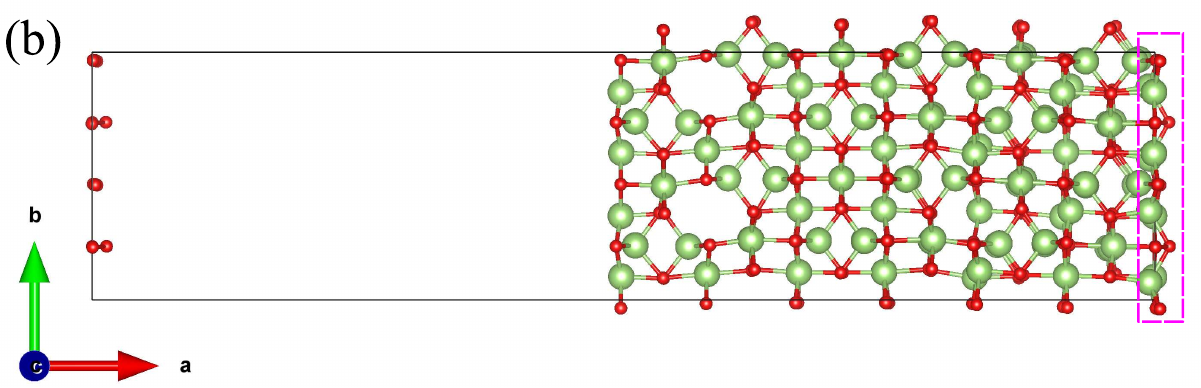}
         \label{fig:terminate-b}
     \end{subfigure}
      \begin{subfigure}{0.49\textwidth}
         \centering
         \includegraphics[width=7.5cm]{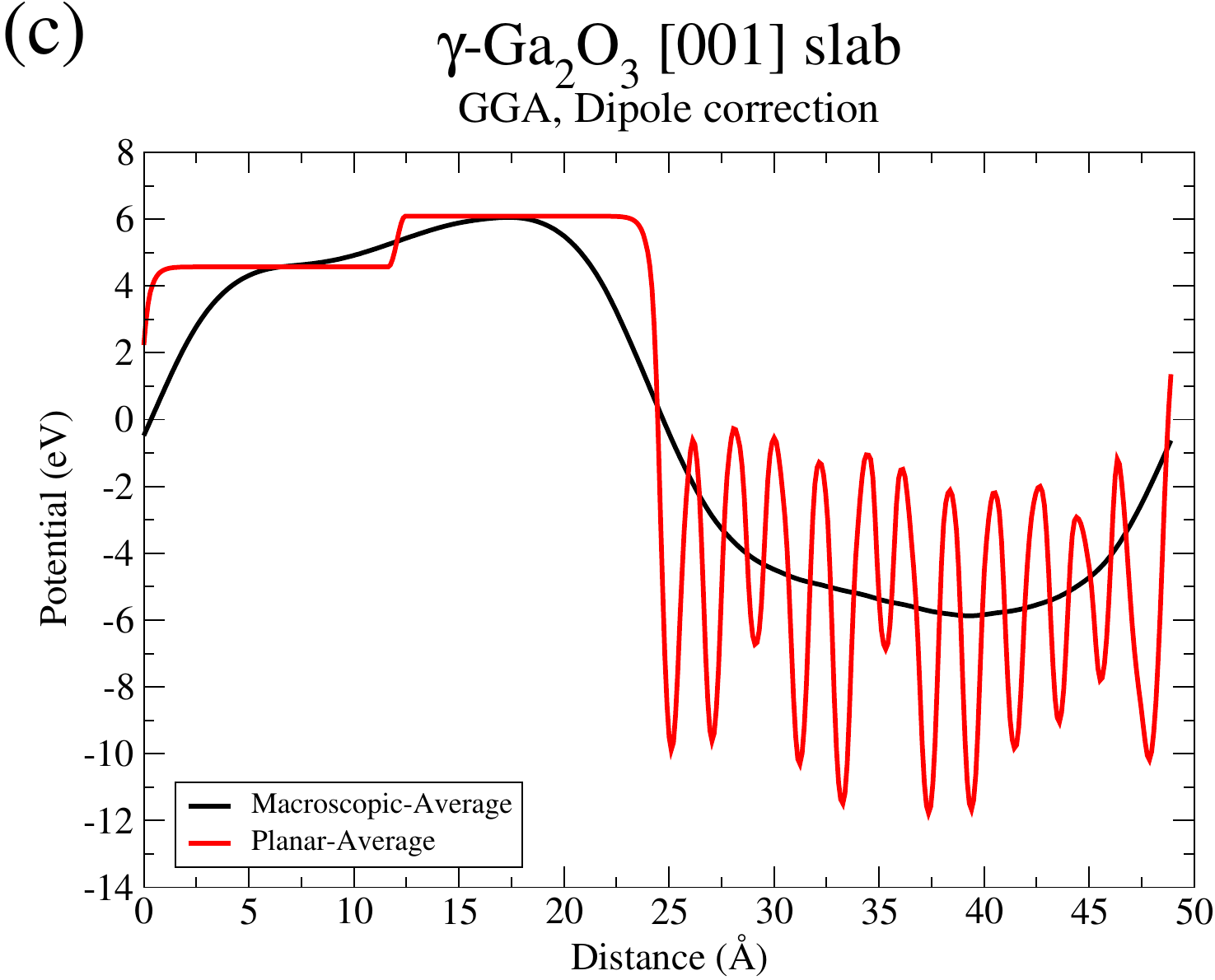}
         \label{fig:terminate-c}
     \end{subfigure}
     \begin{subfigure}{0.49\textwidth}
         \centering
         \includegraphics[width=7.5cm]{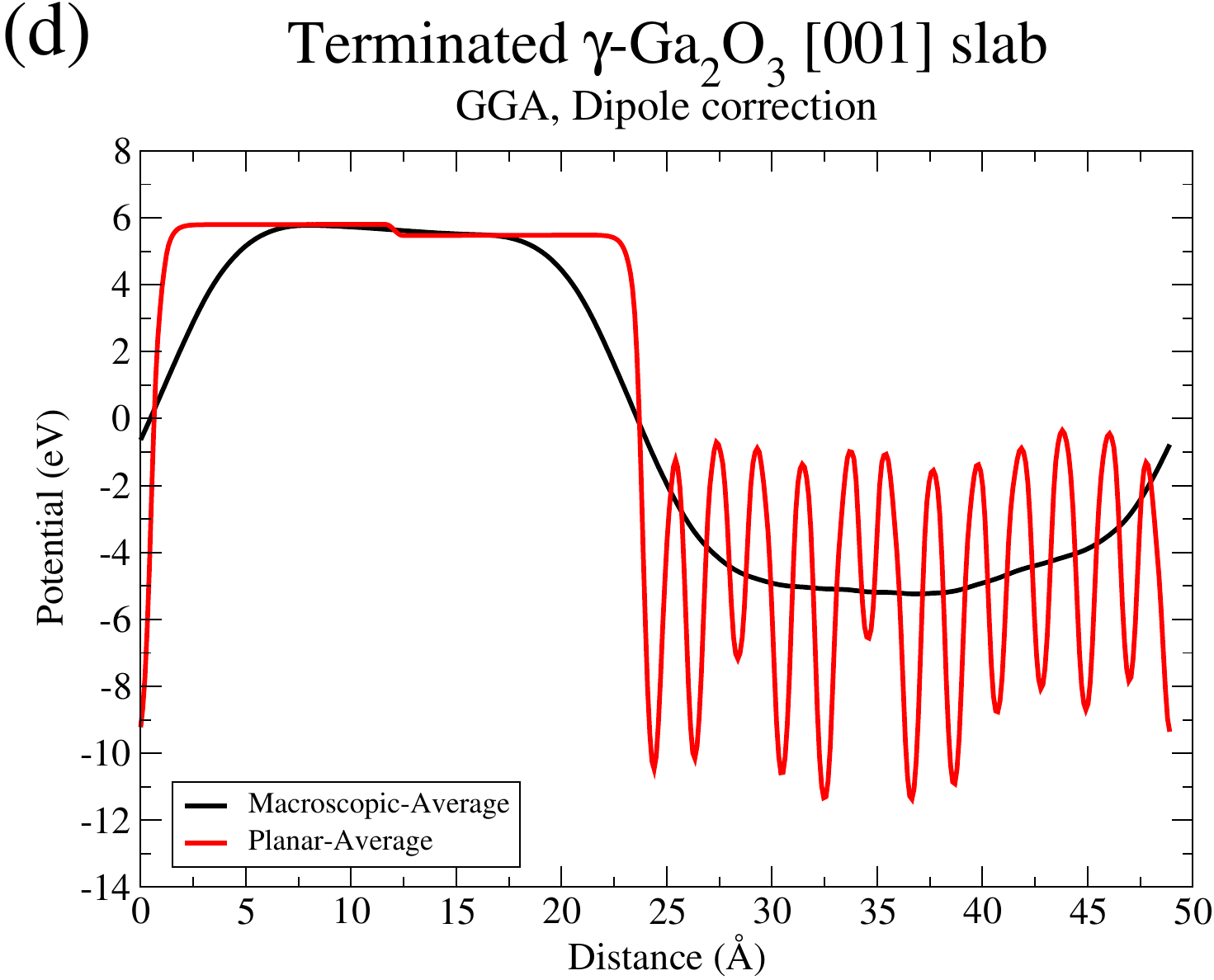}
         \label{fig:terminate-d}
     \end{subfigure}    
     \begin{subfigure}{0.49\textwidth}
        \centering
         \includegraphics[width=8.6cm]{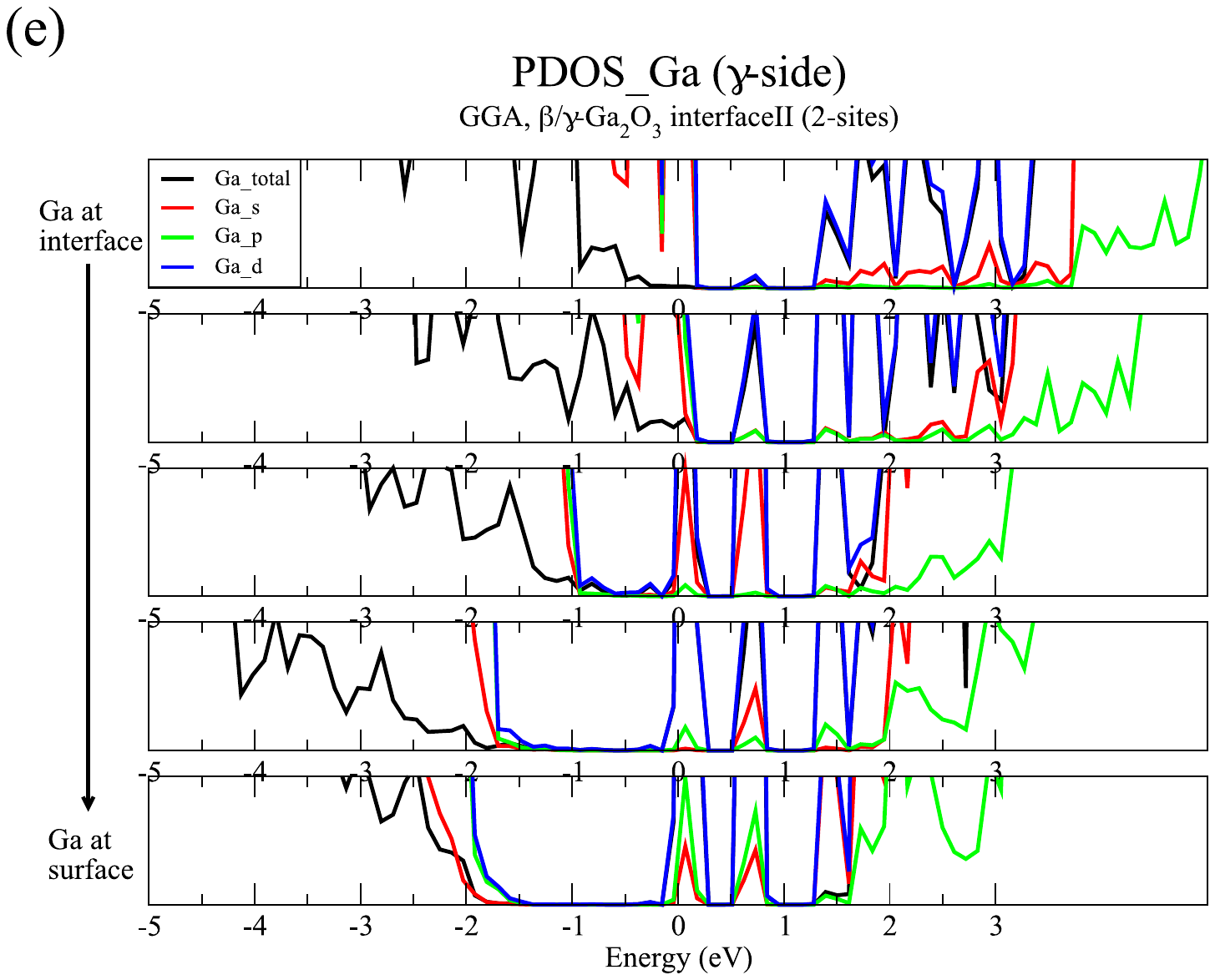}
         \label{fig:terminate-e}
     \end{subfigure}  
     \begin{subfigure}{0.49\textwidth}
        \centering
         \includegraphics[width=8.6cm]{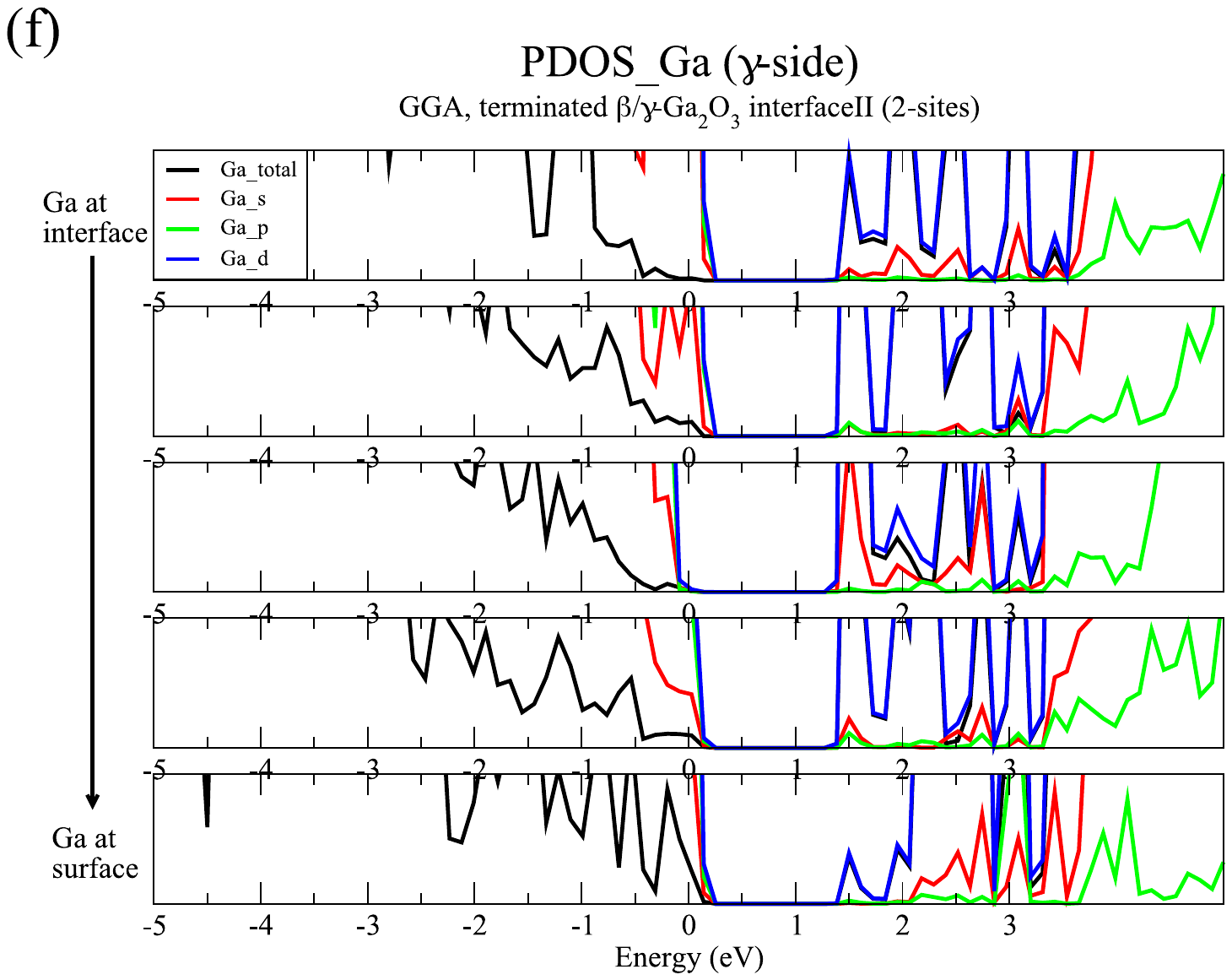}
         \label{fig:terminate-f}
     \end{subfigure}  
     
        \caption{Illustrations of 2-site $\gamma$-\ce{Ga2O3} $\langle$001$\rangle$ (a) stoichiometric slab and (b) octa-Ga terminated slab. Macroscopic-averaged and planar-averaged potentials of the (c) stoichiometric slab and (d) the octa-Ga terminated slab. Projected density of states of Ga ions at the $\gamma$-phase side of (e) stoichiometric $\beta / \gamma$-\ce{Ga2O3} interface \Romannum{2} (2-site) and (f) octa-Ga terminated interface \Romannum{2}.}
        \label{fig:terminate}
\end{figure*}

Fig.~\ref{fig:terminate} illustrates the potential energy distribution within the fully relaxed crystal structures of the [001]-$\gamma$  slab without and with the octa-Ga surface termination as well as the corresponding averaged potentials of these two slabs. 
As shown in Fig.~\ref{fig:terminate}(a) and (b), without the octa-Ga surface termination, Ga cations at the surface move away from the vacuum as a result of the formation of Ga-Ga bonds, while Ga-sublattice is more stable when supported by the extra Ga$_\mathrm{Oct}$ layer.
In addition, the induced by structure distortion electric dipole introduces the presence of a built-in electric field.
This polarization results in charge transfer which may be exaggerated in comparison with the experiment. 
The suggested octa-Ga surface termination can compensate for this polarization sufficiently as shown in Fig.~\ref{fig:terminate}(c) and (d). 

Fig.~\ref{fig:terminate}(e) shows the PDOS of Ga cation in the Interface \Romannum{2} model calculated at the GGA level (where the $\gamma$-cell is the 2-site model).
In each panel, top to bottom are the PDOS of Ga ions from the interface to the open surface of the $\gamma$ side. 
Fig.~\ref{fig:terminate}(f) is the PDOS of the interface without octa-Ga surface termination and (b) is the one terminated by the Ga$_\mathrm{Oct}$ 
layer. Interface states governed by the \textit{d}-orbital electrons are observed in Fig.~\ref{fig:terminate}(e) making the interface nearly metallic.
The terminated system exhibits the presence of a $E_\mathrm{g} = 1.13$ eV which is consistent with the bulk calculation result $1.19$ eV.

\section*{Acknowledgements}
We acknowledge the M-ERA.NET Program for financial support via the GOFIB project administrated in Finland by the Academy of Finland project number 352518 and by the Research Council of Norway project number 337624. International collaboration was partially supported by the INTPART Program at the Research Council of Norway, project number 322382. Additional support was received from the Research Council of Norway, project number 351033. 
J. Zhao acknowledges the National Natural Science Foundation of China under Grant 62304097; Guangdong Basic and Applied Basic Research Foundation under Grant 2023A1515012048; Shenzhen Fundamental Research Program under Grant JCYJ20230807093609019.
The CSC -- IT Center for Science, Finland, is acknowledged for computational resources. 

\bibliographystyle{apsrev4-2}

\bibliography{bandoffset}

\end{document}